\documentclass[aps,preprint]{revtex4}
\usepackage{amsfonts}
\usepackage{amssymb,epsfig}
\usepackage{latexsym}
\usepackage{amsmath}
\usepackage{graphicx}

\begin{document}

\title{Best values of parameters for interacting HDE with GO
IR-cutoff in Brans-Dicke cosmology }
\author{A.
Khodam-Mohammadi$^{1}$\footnote{khodam@basu.ac.ir},
 E.
 Karimkhani$^{1}$\footnote{E.karimkhani91@basu.ac.ir}
 and A. Sheykhi$^{2}$ \footnote{asheykhi@shirazu.ac.ir}
} \address{$^1$ Department of Physics, Faculty of Science, Bu-Ali
Sina University, Hamedan 65178, Iran\\
$^2$ Physics Department and Biruni Observatory, College of Sciences,
Shiraz University, Shiraz 71454, Iran\\
Research Institute for Astronomy and Astrophysics of Maragha
(RIAAM), P.O. Box 55134-441, Maragha, Iran}

\begin{abstract}
We investigate the interacting holographic dark energy (HDE) with
Granda-Oliveros (GO) IR-cutoff in the framework of Brans-Dicke (BD)
cosmology. We obtain the equation of state (EoS) parameter of HDE,
$w_D$, the effective EoS parameter $w_{\mathrm{eff}}$, the
deceleration parameter $q $ and the squared of sound speed $v_s^2$
in a flat FRW universe. We show that at late time the cosmic
coincidence problem can be alleviated. Also we show that for
non-interacting case, HDE can give a unified dark matter-dark energy
profile in BD cosmology, except that it cannot solve the coincidence
problem in the future. By studying the equation of state parameter,
we see that the phantom divide may be crossed. Using the latest
observational data, we calculate the best values of the parameters
for interacting HDE in BD framework. Computing the deceleration
parameter implies that the transition
from deceleration to the acceleration phase occurred for redshift $z\geq 0.5$%
. Finally, we investigate the sound stability of the model, and find
that HDE with GO cutoff in the framework of BD cosmology can lead to
a stable DE-dominated universe favored by observations, provided we
take $\beta=0.44$ and $b^2<0.35$. This is in contrast to HDE model
in Einstein gravity which does not lead to a stable DE dominated
universe.
\end{abstract}

\maketitle

\address{$^1$ Department of Physics, Faculty of Science, Bu-Ali
Sina University, Hamedan 65178, Iran\\
$^2$ Physics Department and Biruni Observatory, College of
Sciences, Shiraz University, Shiraz 71454, Iran}

%\keywords{holographic dark energy; coincidence problem;
%Granda-Oliveros cutoff}

%\PACS{{95.36.+x}{Dark energy} \and {98.80.-k}{Cosmology}}}

\section{ INTRODUCTION}

Nowadays, countless and precise cosmological observations, type Ia
supernova (SNIa) \cite{RO1998}, the 9th year data of the WMAP
mission \cite{WMAP9}, the cosmic microwave background radiation
\cite{HAN2000} and the large scale structure (LSS) \cite{COL2001},
confirm that our Universe is currently undergoing a phase of
accelerated expansion. The provenance of acceleration may be caused
due to an un-known energy component with negative pressure, called
\textit{dark energy} (DE). According to cautious analysis of
cosmological observations, about $\%73$ of the total energy content
of the universe has been occupied by DE, around $\%23$ pressureless
dark matter (DM), and around $\%4$ of total energy content is
denoted to the normal baryonic matter. However, at the present, the
radiation part can be ignored in comparison to other components.
Despite of mysterious nature of DE, during the past decades, many
candidates have been nominated in order to describe DE. See
\cite{PADMAN2003} and references therein. The first and simplest
candidate of dark energy is the cosmological constant $\Lambda$,
with a constant equation of state (EoS) parameter $w_{\Lambda }=-1$.
Although this model is consistent very well with all observations,
it suffers the cosmic coincidence problem. The solution of the
cosmic coincidence problem requires that our universe behaves in
such a way that the ratio of DM to DE densities must be a constant
of order unity or varies more slowly than the scale factor and
finally reaches to a constant of order unity
\cite{ZPAVON0606555,Bisabr,khodam11}. In order to solve this
problem, the dynamical DE models have been proposed. Some analysis
on the `SNIa' observational data reveals that the time varying DE
model gives a better fit compare with a cosmological constant
\cite{SOLA2005}.

Among of many dynamical DE models, the so called HDE model, based on
holographic principle proposed by 't Hooft \cite{HOOFT9310026} and Susskind
\cite{SUSS1995}, has attracted a lot of attention. According to the
holographic principle, the number of degrees of freedom of a physical system
scales with its area instead of its volume. The development of holographic
principle for our purpose was put forwarded by Cohen et al. \cite{COHEN1999}
specify an infrared cutoff length scale and then by Hsu \cite{HSU2004} and
Li \cite{LI2004}, who applied HDE model for solving the DE puzzle. In this
model the energy density is written by $\rho _{D}=3\eta ^{2}M_{pl}^{2}/L^{2}$%
, where $L^2$ is proportional to the area which provides an IR-cutoff, $%
M_{pl}$ is the Planck mass and the numerical constant $3\eta ^{2}$ is
introduced for convenience and utility \cite{HSU2004,LI2004,MYUNG2007}. At
following, we would work in the framework of \emph{\ natural unit}, where ($%
c=\hbar =1$). The IR-cutoff '$L$' plays an essential role in HDE model. If $L
$ is chosen as particle horizon, the HDE can not produce an acceleration
expansion \cite{hsu}, while for future event horizon, Hubble scale '$L=H^{-1}
$', and apparent horizon (AH), as an IR-cutoff, the HDE can simultaneously
drive accelerated expansion and solve the coincidence problem \cite%
{Pavon1,ZPAVON0606555,Sheykhi1}. Thereupon, Gao et al.
\cite{GAO0712.1394} recommended that the HDE density may inversely
be proportional to the Ricci scalar curvature. Succeeding this, Feng
\cite{FENG2008} studied this model in the framework of BD cosmology.
Afterward on Granda-Oliveros \cite{GO2008,easson2011} proposition, a
new cutoff based on wholly dimensional basis, which adds a term
including the first derivation of the Hubble parameter, was
introduced. This cutoff looked alike the Ricci scalar of the FRW
metric but with two
free parameters, $\rho _{D}=3M_{\mathrm{pl}}^2( \gamma H^{2}+\beta \overset{%
\cdot }{H})$, where $\gamma $\ and $\beta $\ are constant parameters
of order unity. This model depends on local quantities, avoiding in
this way the causality problem that is exist in the holographic dark
energy based on the event horizon \cite{GO2009}. Despite of some
success in corresponding with other DE models like scalar field
\cite{khodam111}, Chaplygin gas \cite{m-kh2011} and study of
cosmological evolution \cite{Sharif, Malek} and obtaining the
cosmological constrain on this model \cite{Li}, this form of cutoff
unable to fit the most recent growth data on structure formation in
the ordinary Einstein general relativity frame work
\cite{bas-pol2012,bas-sola2014}.

For the reason that the HDE density belongs to a dynamical cosmological
constant, we need a dynamical framework to accommodate it instead of
Einstein gravity. The best choice for this, is BD theory which is the scalar
tensor theory and was invented first by Jordan \cite{JORDAN1955} and then
ripened by brans and Dicke \cite{BD1961}. This theory is based on Mach's
principle, which is a fundamental principle to explain the origin of
inertia. In attempting to incorporate Mach's principle, the BD theory
introduces a time dependent inertial scalar field $\varphi $, which plays
the role of the gravitational constant $G$, so that $\varphi (t)\propto 1/G$
and is determined by the matter field distributions. So the gravitational
fields are described by the metric $g_{\mu\nu}$ and the BD scalar field $%
\varphi $, which has the dimension $[\varphi ]=[M]^{2}$. In BD theory, the
scalar field $\varphi $ couples to gravity via a coupling parameter $\omega $
and it has been generalized for various scalar tensor theories. Therefore,
the investigation on the holographic models of DE in the framework of BD
theory, has of great interest and have been accomplished in \cite{GONG2004}.

The combination of BD field and HDE can accommodate $w_{D}=-1$
crossing for the EoS parameter of non-interacting HDE
\cite{SHEYKHI0907.5458}. It was shown that in BD cosmology when an
interaction between DE and DM is taken into account, the transition
from normal state $w_{D}>-1$ to the phantom regime $w_{D}<-1$ can be
more easily accounted compared to the Einstein gravity. Also, in
\cite{XULI2009}, it has been demonstrated that the accelerated
expansion will not be achieved in BD theory without interactions,
when the Hubble horizon is taken as the IR cut-off. Howbeit, when
the event horizon takes the role of IR cut-off, an accelerating
universe is obtained. Furthermore, the phantom crossing is more
easily
achieved when the matter and the HDE undergo an exotic interaction \cite%
{JAMILKARAMI2011}.

In this paper we investigate the HDE in BD cosmology using a GO as
IR-cutoff. We assume the BD field as a power law of the scale factor , $%
\varphi \propto a^{n}$, and as it has mentioned in \cite%
{JAMILKARAMI2011,BP2007} that there is no compelling reason for this choice.
However it has been shown that for small $\mid n\mid $ it leads to
consistent result. Usually in papers which investigate the DE in the BD
cosmology with different IR-cutoff \cite%
{FENG2008,BP2007,ZPAVON0606555,JAMILKARAMI2011}, authors assume that matter
evolves as $\rho _{m}\propto \rho _{m\circ }a^{-3}$\ or some other
assumptions like this for $\rho _{m}$\ and by use of this, they calculated
the EoS parameter of DE. This assumption had been also considered in
interacting case, which has been given more complicated or sometimes wrong
relations for cosmological quantities. In fact by considering interaction
between DE-DM, energy conservation is not valid separately for each
component. Here we are attempting to provide a more generic way, without any
restricted supposition, to calculate the EoS parameter and some other
cosmological parameters of our interest.

It must be noted that, although, there is not any evidence for
existence any direct interaction between DM and DE, no known
symmetries prevent such interaction. On the other hands, such a
choice for interacting term don't make any discordance with
observation as mentioned in Ref. \cite{RONG-GEN}. In addition, the
latest cosmological constraints on the coupled DE models, in which
the quintessence scalar field non-minimally couples to the cold DM,
from the recent Planck measurements has been done \cite{Planck}.

The last main task that must be investigated is the stability analysis of
model by calculating the squared of sound speed ($v_s^2=dp/d\rho$) \cite%
{peebleratra}. The analysis of squared of sound speed could say us about the
perturbation growth and the structure formation at present. However this
quantity does not enough insight to say the model is surely stable, but at
least can show sounds of instability of the model. The sign of $v_s^2$ plays
a crucial role in determining the instability of the background evolution.
If $v_s^2<0$, it means that we have the classical instability of a given
perturbation. In contrast $v_s^2>0$, leaves chance for greeting a stable
universe against perturbations.

This paper is outlined as follows. In section \ref{BD}, a brief review of
the HDE in BD cosmology is given. In section \ref{flat}, we describe the
physical contest which we are working in and we derive the EoS parameter of
DE, $w_{D}$ , effective EoS parameter ($w_{\mathrm{eff}}$) and the
deceleration parameter $q$ in a flat FRW universe, for various choice of
interaction. The quantity of interest for analyzing the coincidence problem
is the ratio $u= \rho _{m}/\rho _{D}$ , which we make a spacious discussion
on this quantity in this section. In section \ref{sound}, the stability
analysis through the dynamical DE model is studied. In section \ref%
{discussion}, we give a detailed discussion on all calculations. We
summarize our results in section \ref{sum}.

\section{general formalism}

\label{BD}

We start with a brief review on HDE in the framework of BD
cosmology. The BD field equation can be written as \cite{BD1961,
Wein-book}
\begin{eqnarray}
R_{\mu \nu }-\frac{1}{2}g_{\mu \nu }R-\frac{\omega }{\varphi ^{2}}(\nabla
^{\mu }\varphi \nabla _{\nu }\varphi -\frac{1}{2}g_{\mu \nu }\nabla _{\alpha
}\varphi \nabla ^{\alpha }\varphi ) -\frac{1}{\varphi}(\nabla _{\mu }\nabla
_{\nu }\varphi -g_{\mu \nu }\nabla _{\alpha }\nabla ^{\alpha }\varphi )=
\frac{1}{\varphi}T_{\mu \nu },  \label{0-1}
\end{eqnarray}
where $\varphi $ is the BD scalar field which is allowed to vary with space
and time and $\omega $ is the generic dimensionless parameter of the BD
theory. In this theory, the total energy momentum tensor $T_{\mu
}^{\nu}=diag(-\rho,p,p,p)$, minimally couples to the gravity and there is no
interaction between the scalar field $\varphi $ and the matter field.
General relativity is a particular case of the BD theory, corresponding to $%
\omega \rightarrow \infty $ \cite{Wein-book}. In a FRW universe,
with line element metric
\begin{equation}
ds^{2}=-dt^{2}+a^{2}(t)\left[ \frac{dr^{2}}{1-kr^{2}}+r^{2}\left( d\theta
^{2}+\sin ^{2}\theta d\varphi ^{2}\right) \right],  \label{0-5}
\end{equation}
the BD field equations take the form \cite{pavon2001}
\begin{eqnarray}
3\left(\frac{\overset{\cdot }{a}^{2}}{a^{2}}+\frac{k}{a^{2}}\right)-\frac{1}{%
2}\omega \frac{\overset{\cdot }{\varphi }^{2}}{ \varphi ^{2}}+3\frac{\dot {a}%
}{a} \frac{\dot{\varphi}}{\varphi }&=&\frac{1}{\varphi } (\rho _{m}+\rho
_{D})  \label{0-6} \\
2\frac{\ddot{a}}{a}+\frac{\overset{\cdot
}{a}^{2}}{a^{2}}+\frac{k}{a^{2}}+ \frac{1}{2}\omega
\frac{\overset{\cdot }{\varphi }^{2}}{ \varphi ^{2}}+2
\frac{\overset{\cdot }{a}}{a}\frac{\overset{\cdot }{\varphi
}}{\varphi }+ \frac{\overset{\cdot \cdot }{\varphi }}{\varphi }&=&-
\frac{1}{\varphi } p_{D},  \label{0-7}
\end{eqnarray}
where $a(t)$ is the dimensionless scale factor of the universe, $\rho = \rho
_{m}+\rho _{D}$ is the total energy density and the curvature parameter $k =
-1, 0, 1$, represent spatially open, flat and closed universe, respectively.
Also for simplicity we assuming $\varphi =\varphi (t)$. The equation of
motion for BD scalar field is given by $(2\omega +3)\nabla _{\mu }\nabla
^{\mu }\varphi =T$, where $T$ is trace of energy momentum tensor $T_{\mu \nu}
$. Since the DM is pressureless, thus the total pressure equals the DE
pressure, $p=p_{D}$. We further assume that both components do not conserve
separately but interact with each other in such a manner that the continuity
equations take the form
\begin{equation}
\dot{\rho}_{D}+3H(1+w_{D})\rho _{D}=-Q,  \label{0-8}
\end{equation}
\begin{equation}
\dot{\rho}_{m}+3H\rho _{m}=Q,  \label{0-9}
\end{equation}
where $w_{D}=p_{D}/\rho _{D}$, denotes the EoS of DE, and $Q$ stands for the
interaction term. It should be noted that the ideal interaction term must be
motivated from the theory of quantum gravity. In the absence of such a
theory, we rely on pure dimensional basis for choosing an interaction $Q$.
It is worth noting that the continuity equations imply that the interaction
term should be a function of a quantity with units of inverse of time (a
first and natural choice can be the Hubble factor $H$) multiplied with the
energy density. Therefore, the interaction term could be in any of the
following forms: (i) $Q\propto H \rho_D$, (ii) $Q\propto H \rho_m$, or (iii)
$Q\propto H (\rho_m+\rho_D)$. Thus hereafter we consider only the first
case, namely $Q=b^2 H \rho_D=\Gamma \rho _{D}$, where $b^2$ is a coupling
constant.

\section{HDE WITH GO CUTOFF IN FLAT BD theory}

\label{flat}

\label{111} We assume the energy density of HDE with GO cutoff in BD theory
in the form
\begin{equation}
\rho _{D}=3\varphi \left( \gamma H^{2}+\beta \dot{H}\right),  \label{1-1}
\end{equation}
where $M_{\mathrm{pl}}^2$ has been replaced by $\varphi$ in BD cosmology as
mentioned previously. Assume the BD field is proportional to the scale
factor as, $\varphi \propto a^{n}$, it then follows that
\begin{equation}
\frac{\dot{\varphi }}{\varphi }=nH,\ \ \ \frac{\ddot{ \varphi }}{\varphi }
=n^{2}H^{2}+n\dot{H},\ \ \ \frac{\ddot{\varphi }}{\dot{\varphi }}=(n+\frac{
\dot{H}}{H^{2}})H.  \label{1-2}
\end{equation}
Using relations (\ref{1-2}), Eqs. (\ref{0-6}) and (\ref{0-7}) for flat FRW
universe reduce to
\begin{eqnarray}
\rho _{D}&=&\frac{\varphi H^{2}}{(1+u)}\left[ 3(1+n)-\frac{\omega n^{2}}{2} %
\right],  \label{1-3} \\
\rho_{D}&=&-\frac{H^{2}\varphi}{w_{D}}\left[3+\frac{\dot{{H}}}{H^{2}}
(2+n)+n^{2}+\frac{\omega n^{2}}{2}+2n\right],  \label{1-4}
\end{eqnarray}
where $u=\rho _{m}/\rho _{D}$ is the ratio of energy densities.
Differentiating Eq. (\ref{1-1}) with respect to the cosmic time $t$, gives
\begin{equation}
\dot{\rho }_{D}=3\dot{\varphi }\left( \gamma H^{2}+\beta \dot{H}\right)
+3\varphi \left( 2\gamma \dot{H}H+\beta \overset{\cdot \cdot }{H}\right).
\label{1-5}
\end{equation}
Equating Eqs. (\ref{1-3}) and (\ref{1-4}), leads to
\begin{equation}
\frac{\dot{H}}{H^{2}}=-\frac{3w_{D}[(1+n)-\frac{\omega n^{2}}{6}]}{(2+n)(1+u)%
}-\frac{3+n^{2}+2n+\frac{\omega n^{2}}{2}}{(2+n)}.  \label{1-6}
\end{equation}
On the other hand, by inserting Eq. (\ref{1-1}) in (\ref{1-3}), we  arrive
at
\begin{equation}
\frac{\dot{H}}{H^{2}}=\frac{(1+n)-\frac{\omega n^{2}}{6}}{(1+u)\beta }-\frac{
\gamma }{\beta }.  \label{1-7}
\end{equation}

It's useful here to examine the signature of the deceleration parameter, $%
q=- \ddot{a}/aH^{2}$. Since $H=\dot{a}/a$, we have $\ddot{a}/a=\dot{H}+H^{2}$
and from (\ref{1-7}), it follows that
\begin{equation}
q=-1-\frac{\dot{H}}{H^{2}}=-1+\frac{\gamma }{\beta }- \frac{(1+n)-\frac{%
\omega n^{2}}{6}}{(1+u)\beta } .  \label{1-10}
\end{equation}%
Despite of the obtained equation for deceleration parameter in \cite{BP2007}%
, here we see that the BD field, $\varphi $, doesn't appear in Eq. (\ref%
{1-10}) and therefore it caused to make the best estimate for $q$
and $u$ parameters. Recent constrain on interacting DE models in BD
cosmology, gives the best value of $n\approx 0.005$ and $\omega
\approx 1000$ \cite{XULU2011,LU-liu2012}. Because of very small
value of $n$, only terms of order $\omega n$ and $\omega n^{2}$ gets
important and we can neglect again terms of order $n$ and $n^2$ in
calculations. At present time, $u\approx 0.4$, the deceleration
parameter $q$ can obviously be negative if
\begin{equation}
\omega n^{2}<6-8.4\beta \left(\frac{\gamma }{\beta }-1\right),  \label{1-15}
\end{equation}%
where it can give a bound on $\omega n^{2}$ at present by obtaining $\beta $
and $\gamma $. By equating Eqs.(\ref{1-6}) and (\ref{1-7}), the EoS
parameter $w_{D}$, of the HDE in BD theory is given by
\begin{equation}
w_{D}=\frac{1}{3}\left[ A(1+u)-\frac{2+n}{\beta }\right]  \label{1-8}
\end{equation}%
where
\begin{equation}
A=-\frac{1}{1+n-\frac{\omega n^{2}}{6}}\left[ 3+n^{2}+2n+\frac{\omega n^{2}}{%
2}-\frac{(2+n)\gamma }{\beta }\right] .  \label{1-8A}
\end{equation}%
Neglecting terms of order $n\approx 0.005$ while keeping $\omega
n^{2}\approx 0.025 $, we find
\begin{equation}
w_{D}=-\frac{2}{3\beta }-\left( 1+u\right)\left(1-\frac{4\frac{\gamma }{%
\beta }-2\omega n^{2}}{6-\omega n^{2}}\right).  \label{1-9}
\end{equation}%
Also by considering Eq. (\ref{1-6}) and solving it for $w_{D}$ in term of $q$
($q=-1-\dot{H}/H^{2}$), we obtain
\begin{eqnarray}
w_{D}&=&-\frac{(1+u)}{3}\left( \frac{1+n+n^{2}+\frac{\omega n^{2}}{2}-q(2+n)%
}{1+n-\frac{\omega n^{2}}{6}}\right)  \notag \\
&\approx&-(1+u)\left( \frac{2+\omega n^{2}-4q}{6-\omega n^{2}}\right),
\label{1-11}
\end{eqnarray}
where in the last step, we have neglected again terms of order $n$
and $n^2$ and keep only terms of order $\omega n\approx 5.0$ and
$\omega n^{2}\approx 0.025 $. Taking $u\approx 0.4$ and $ q=-0.6$
for the present time, which has been parameterized recently by Pav
\'{o}n et al. \cite{PD2012}, and using Eq (\ref{1-11}), we could
estimate the EoS parameter, $w_{D0}\simeq -1.04$, which is
consistent with observational data of \cite{SUKUZI2012,PLANCK} and
WMAP9+SNLS+HST data \cite {XIALI2013}.

The effective EoS parameter comes out to be
\begin{equation}
w_{\mathrm{eff}}=\frac{p}{\rho }=\frac{w_{D}}{1+u}=-\left( \frac{2+\omega
n^{2}-4q}{6-\omega n^{2}}\right) .  \label{1-11-eff}
\end{equation}%
It is worth while to mention that the acceleration ($q\leq 0$) in BD
cosmology started at $w_{\mathrm{eff}}\leq -(2+\omega n^{2})/(6+\omega
n^{2})\approx -0.34$ which has a very small difference from $-1/3$. This is
only due to $\omega n^{2}$ term. The quantity $w_{\mathrm{eff}}$ can also
determine whether Universe evolves in phantom phase or not. Besides, by
considering super acceleration phase, where $\dot{H}>0$ (i.e. $q<-1)$, the
effective EoS parameter (\ref{1-11-eff}) reduced to $w_{\mathrm{eff}}<-1$,
which coincides perfectly with phantom phase.

By taking the time Derivative of $u$, and using Eqs. (\ref{0-8}) and (\ref%
{0-9}), we can obtain
\begin{equation}
\dot{u}=3Hu\left[ w_{D}+\frac{b^{2}}{3}\left( \frac{1+u}{u}\right) \right] .
\label{1-12}
\end{equation}%
If we now define the e-folding $x$ with definition $x=\ln a=-\ln(1+z)$,
where $z =a^{-1}-1$, is the redshift parameter and using the fact that $%
d/d(x)=\frac{1}{H}d/d(t)$, then Eq. (\ref{1-12}) yields
\begin{equation}
u^{\prime }=u\left[ A(1+u)-\frac{2+n}{3\beta }\right] +b^{2}\left( 1+u\right)
\label{1-13}
\end{equation}%
where prime denotes derivative with respect to $x$ and $\Gamma /H=b^{2}$ is
the interaction parameter. By solving the differential equation (\ref{1-13}%
), the ratio of energy densities, $u(x)$, would be gained as
\begin{eqnarray}
u(x)=\frac{1}{2\beta A}\left\{ C\tan \left[ \frac{Cx }{2\beta }+\arctan
\left( \frac{9\beta A-5n+5\beta b^{2}-10)}{5C}\right) \right]-\beta
A+2+n-\beta b^{2} \right\}  \label{1-14}
\end{eqnarray}%
where the parameter $C$ is given by
\begin{equation}
C=\sqrt{4A\beta (n+2)-\left( \beta b^{2}-2-A\beta -n\right) ^{2}}.
\end{equation}
In limiting case of ordinary Einstein general relativity, (where
$n\rightarrow0,~\omega n^{2}\rightarrow 0,~\omega\rightarrow \infty
$), we find
\begin{equation}
w_{D}=-\left[ (1-\frac{2\gamma }{3\beta })(1+u) +\frac{2}{3\beta }\right].
\label{1-16}
\end{equation}
For a DE dominated universe, where $u=0$, it reduces to the EoS
parameter of Granda-Oliveros \cite{GO2008}
\begin{equation}
w_{D}=- 1-\frac{2}{3\beta }\left( 1-\gamma \right).  \label{1-17}
\end{equation}
However, a comparison between Eqs. (\ref{1-16}) and (\ref{1-9}) shows that
crossing the phantom divide line for the HDE in BD gravity can be more
easily achieved for than when resort to the Einstein gravity. The effective
EoS parameter with above limiting case becomes
\begin{equation}
w_{\mathrm{eff}}=\frac{p}{\rho }=\frac{w_{D}}{1+u}=-\left( \frac{1-2q}{3}%
\right).
\end{equation}
By taking $q=-0.6$ \cite{PD2012}, we have $w_{\mathrm{eff}}=-0.733$, which
shows the quintessence phase of the universe for present time. In section %
\ref{discussion}, we will discuss on all obtained quantities.

Using the continuity equation (\ref{0-9}), the energy density of
dark matter in the interacting case yields
\begin{equation}
\frac{\rho _{m}}{\rho _{m_0}}=\exp
[-3x+3b^{2}(\mathcal{F}(x)-\mathcal{F}(x_0))]
\end{equation}
where $\rho _{m_0}$ is current value of matter energy density and
\begin{eqnarray}
&&\mathcal{F}(x)=\int u(x)dx=\frac{1}{2A}\times\notag\\
&&\Big{\{}x(-A- b^{2}+\frac{1}{\beta })+ \ln \left( 1+\tan \left[
\frac{C x}{2\beta }+\arctan \left( \frac{9\beta A-5n+5\beta
b^{2}-10}{5C}\right) \right] ^{2}\right)\Big{\}}
\end{eqnarray}
and $\mathcal{F}(x_0)$ is the current value of $\mathcal{F}(x)$. In
comparison to the standard matter density law $\rho_{m_s}\propto
exp(-3x)=a^{-3}$, it shows that
$\rho_{m}/\rho_{m_s}=exp[3b^2(\mathcal{F}(x)-\mathcal{F}(x_0))]$,
which is related to interaction parameter, BD parameter and $n$.
This departure of standard law is small. More discussion will be
left by Sec. \ref{discussion}. It is worth wile to mention that for
non-interacting case, $b^2=0$, the matter energy density can be
given by the standard law. The dark energy density is also given by
$\rho_D=u(x)\rho_m$.

The evolution of gravitational constant $G\propto 1/\varphi$ by
considering $\varphi\propto a^n$, is obtained as
$|\dot{G}/G|=\dot{\varphi}/\varphi=n H$. The upper bound of this
quantity with given value $n=0.005$, has a good agreement with
latest constraint on the evolution of G at present which is
$|\dot{G}/G|< 10^{-11}yr^{-1}$ \cite{ray2007,XULU2011}. Also if we
consider the another form of evolution of $\Delta G/G_0\sim -n
ln(a)$ or $\dot{G}/G_0\sim -n H$, which is suggested in some
theories(e.g. in fact theoretical QFT models
\cite{j-sola2008,bas-pol2012,grande-sola2008}), ($G_0$ is current
value of $G$), very slow evolution of $G$ is also expected by our
given value of $n=0.005$.

\section{Sound Stability of the model}

\label{sound} From observations we know that our universe is in a DE
dominated phase. Thus any viable DE model should result a stable DE
dominated universe. One simple way to check such a stability for any new DE
model is to discuss the behavior of the square sound speed ($v_s^2 = dp/d\rho
$) in a DE dominated universe \cite{peebleratra}. The sign of $v_s^2$ plays
a crucial role in determining the stability of the background evolution. If $%
v_s^2<0$, it means that we have the classical instability of a given
perturbation. In contrast $v_s^2>0$, leaves chance for greeting a stable
universe against perturbations. However, this does not enough insight to say
the model is surely stable but at least can show sounds of instability in
the model. This approach has been used for exploring some DE models. For
example in \cite{myung1,myung2} the authors investigated the behavior of the
square sound speed for HDE as well as the agegraphic DE models and found
both of these models are instable against background perturbations. Also it
was shown that chaplygin gas and tachyon DE have positive squared speeds of
sound with, $v_s^2= -w$, and thus they are supposed to be stable against
small perturbations \cite{gorini,sandvik}. The stability of the GDE models
was studied in \cite{ebrins}, and it was shown that the GDE models are not
capable to result a stable DE dominated universe. Also, a same procedure was
considered in \cite{saaidi} to show the stability of the GDE in the
chameleon BD theory.

In our model, $\rho =\rho _{D}(1+u)$ and  for pressureless CDM,
$p=p_D$, the quantity $v_{s}^{2}$ for a flat FRW universe is
obtained as
\begin{equation}
v_{s}^{2}=\frac{\overset{\cdot }{p}}{\overset{\cdot }{\rho }}=\frac{\overset{%
\cdot }{\rho _{D}}w_{D}+\rho _{D}\overset{\cdot }{w_{D}}}{\overset{\cdot }{%
\rho _{D}}(1+u)+\rho _{D}\overset{\cdot }{u}}  \label{3-1}
\end{equation}%
Taking the time derivative of Eq. (\ref{1-8}) and using Eqs.
(\ref{0-8}) and (\ref{1-12}), we
can obtain the following equation for sound speed with respect of $x$%
\begin{equation}
v_{s}^{2}=-\frac{w_{D}\left( Au(x)-3w_{D}-b^{2}-3\right) +\frac{b^{2}}{3}%
A\left( 1+u(x)\right) }{3(1+u(x)+w_{D})}.  \label{2-1}
\end{equation}

Detailed discussion will be given in the next section.

\section{discussion}\label{discussion}

In this work for simplicity and in order to reduce the unconstrained
parameters, we choose $\gamma =2\beta $. It is a reasonable
assumption since in Refs. \cite{WANGXU2010,GO2008} the numerical
values for these parameters has been obtained by use of
observational data in Einstein gravity, which confirm that $\gamma
\approx 2\beta $. Also this assumption is very accurate in Ricci DE
model \cite{khodam111}.

We start our analysis by computing $u(x)$ with respect to $x$ and then the
behavior of all cosmological quantities such as the EoS of DE, $w_{D}$,
effective equation of state, $w_{\mathrm{eff}}$, deceleration parameter, $q$%
, and $v_{s}^{2}$ are discussed. At present time where $x=0$
($z=0$), the following values of cosmological parameters
``$u=0.4,~q=-0.6,~\omega n^{2}=0.025,~\mid n\mid \approx 0$" are
assumed, which have been chosen as the best values of recent
observational data \cite{XULU2011, PD2012}.

\begin{figure}[tbp]
\epsfxsize=7cm\centerline{\epsffile{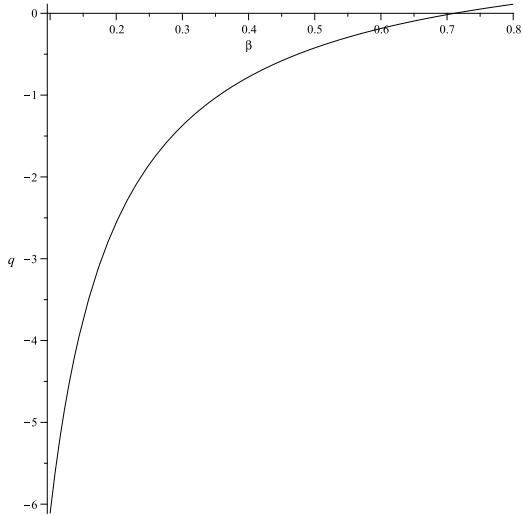}}
\caption{behavior of $q$ versus $\protect\beta $ at present ($x=0$)}
\label{fig1}
\end{figure}

First, we discuss on the deceleration parameter $q$ in the spatially flat
FRW case. From the mentioned cosmological data and Eq. (\ref{1-10}), the
parameter $\beta $ is calculated as $\beta \approx 0.44$. As it is shown in
Fig. \ref{fig1}, in order to have the present acceleration expansion ($q<0$%
), an upper limit $\beta <0.71$ is found and from Eq. (\ref{1-7}),
for supper acceleration expansion ($\dot{H}>0$), this limit reduced
to $\beta <0.36$. Therefore, it shows that Ricci DE model which is a
famous model of
HDE with Ricci scaler as IR cutoff, namely $\rho _{D}=3C^{2}\varphi (2H^{2}+%
\dot{H})$, the acceleration expansion can be achieved in BD cosmology
provided $C^{2}<0.71$.

\begin{figure}[tbp]
\epsfxsize=7cm\centerline{\epsffile{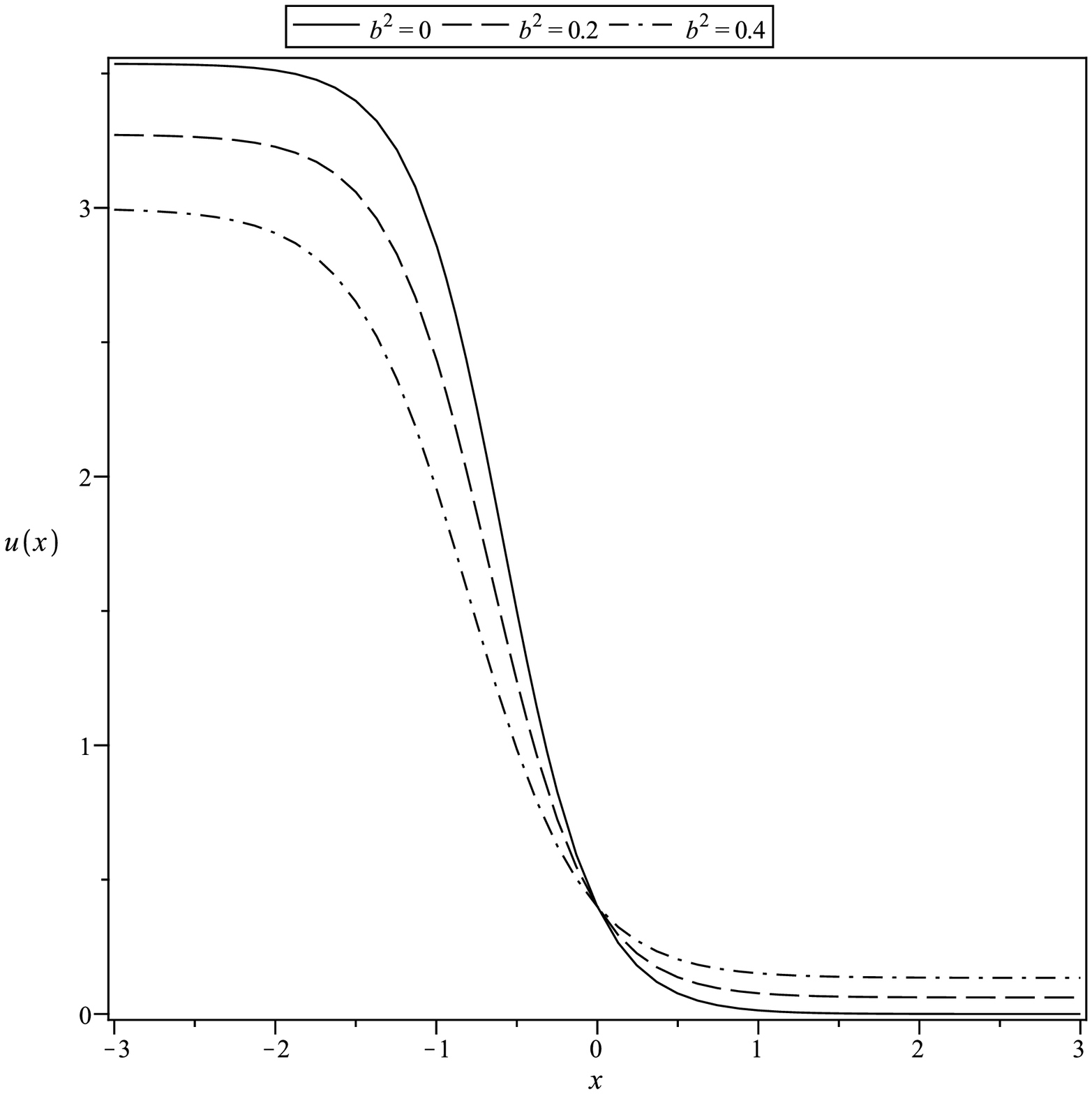}} \caption{evolution of
$u(x)$ versus $x$ for various $b^{2}$} \label{fig2}
\end{figure}

\begin{figure}[tbp]
\epsfxsize=7cm\centerline{\epsffile{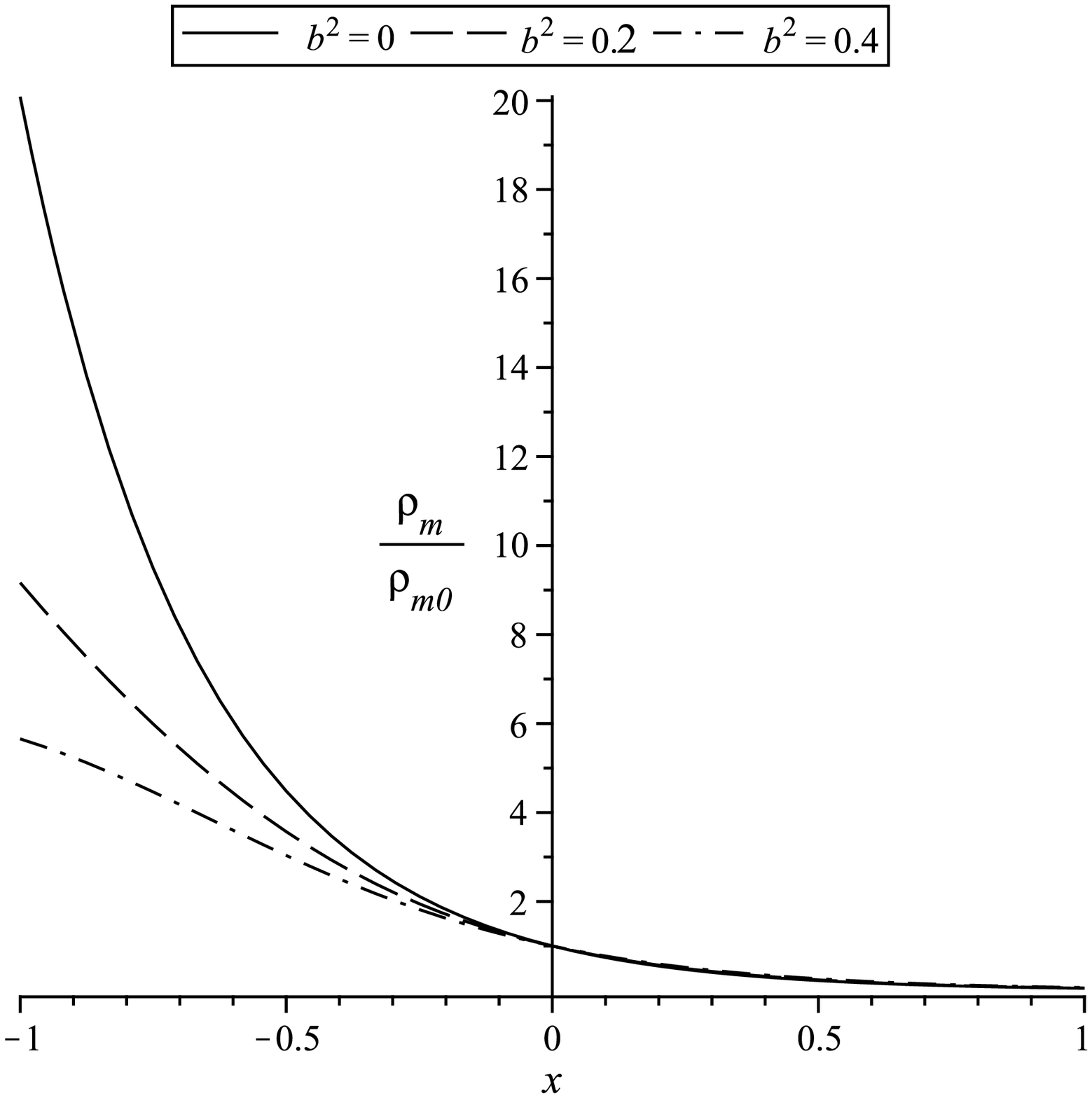}} \caption{evolution of
$\rho_m/\rho_{m_0}$ versus $x$ for various $b^{2}$} \label{fig2-1}
\end{figure}
\begin{figure}[tbp]
\epsfxsize=7cm\centerline{\epsffile{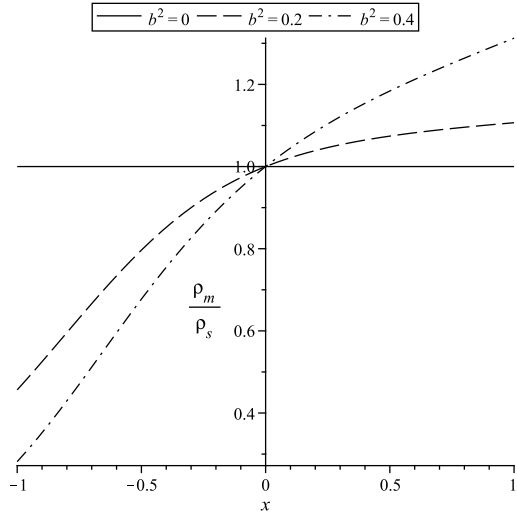}} \caption{evolution of
$\rho_m/\rho_{m_s}$ versus $x$ for various $b^{2}$} \label{fig2-2}
\end{figure}
\begin{figure}[tbp]
\epsfxsize=7cm\centerline{\epsffile{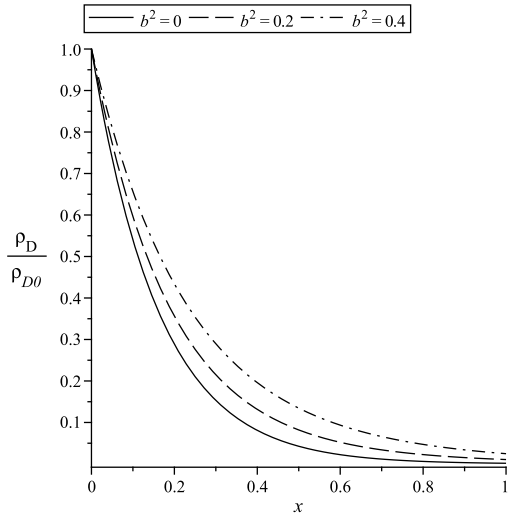}} \caption{evolution of
$\rho_D/\rho_{D0}$ versus $x$ for various $b^{2}$} \label{fig2-3}
\end{figure}

In Fig. \ref{fig2}, the evolution of $u(x)$ is plotted versus $x$
for various interacting parameters. It shows that $u(x)$ is bounded
which depend on interaction parameter. By increasing the interacting
parameter, $u(x)$ evolves slower and will reach to a finite value
according to interacting parameter at infinity and at past the upper
limit of $u(x)$ is of order unity and it reached to a smaller
saturated value, for larger values of $b^2$. For example $u(x)$ is
bounded between (3.00-0.13) for $b^2=0.4$ and (3.27-0.06) for
$b^2=0.2$. The same of this behavior is seen in Refs.
\cite{gra-sola2006,gra-sola2009}. For non-interacting case $u(x)$
vanishes at late time. Therefore it can alleviate the coincidence
problem only for interacting model. Behavior of matter energy
density versus $x$ for various $b^2$, is illustrated in Fig.
\ref{fig2-1} and Fig. \ref{fig2-2} shows the ratio of matter energy
density with standard one in versus $x$. These figures show that the
departure of matter energy density with respect to standard law (or
non-interacting case) become smaller at future (see figure
\ref{fig2-1}) and it become larger by increasing the interaction
term (see figure \ref{fig2-2}). The evolution of dark energy density
is plotted in Fig. \ref{fig2-3}. As it is shown, $\rho_D$ reduce to
a finite non zero value for interacting DE models.

\begin{figure}[tbp]
\epsfxsize=7cm\centerline{\epsffile{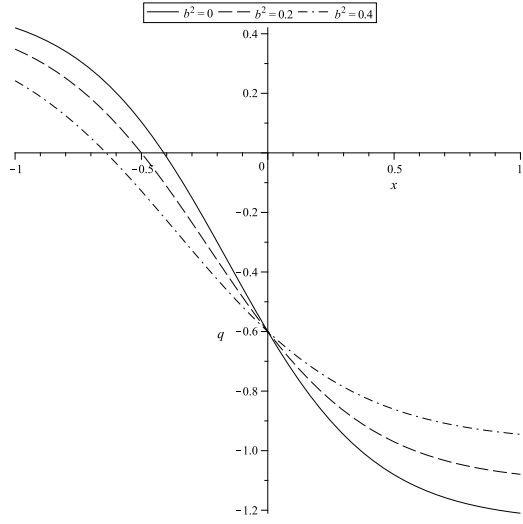}} \caption{Evolution of
$q$ versus $x$ for various $b^2$} \label{fig3}
\end{figure}

Inserting $u(x)$ from (\ref{1-14}) into (\ref{1-10}), the deceleration
parameter, $q$ is calculated with respect to $x$ for the best value of $%
\beta\approx 0.44$. Figure \ref{fig3} shows that for all values of $b^2$,
the deceleration parameter transits from deceleration ($q>0$) to
acceleration ($q<0$) at $x<-0.4$ which corresponds to $~z>0.5$. However by
increasing $b^2 $ the transition point moves to older Universe. Further it
shows that the interaction will affect on late time acceleration. Increasing
$b^2$ corresponds with decreasing $q$ at any time.

\begin{figure}[tbp]
\epsfxsize=7cm\centerline{\epsffile{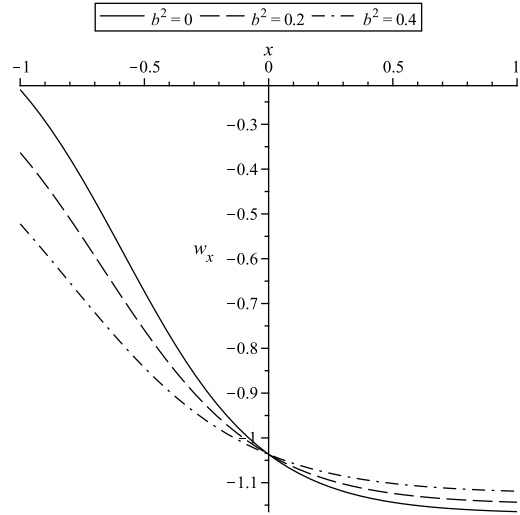}} \caption{Evolution of
$w_{D}$ versus $x$ for various $b^{2}$} \label{fig4}
\end{figure}

\begin{figure}[tbp]
\epsfxsize=7cm\centerline{\epsffile{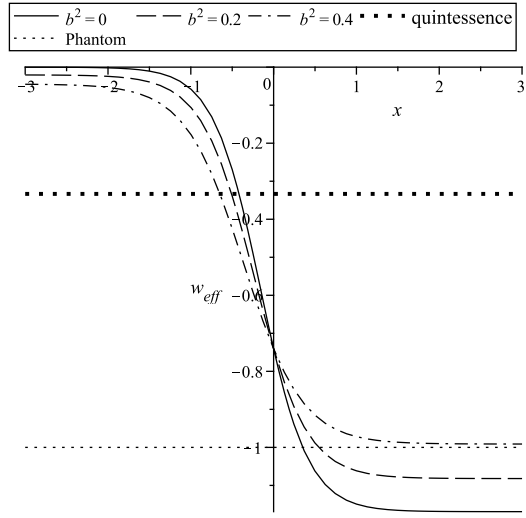}} \caption{Evolution of
$w_{\mathrm{eff}}$ versus $x$ for various $b^{2}$} \label{fig5}
\end{figure}

\begin{figure}[tbp]
\epsfxsize=7cm\centerline{\epsffile{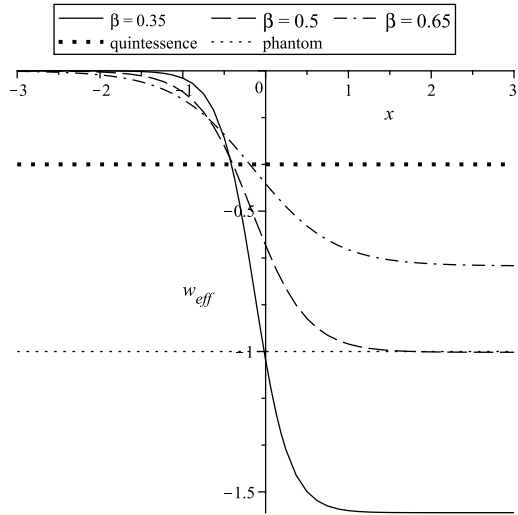}} \caption{behavior of
$w_{\mathrm{eff}}$ versus $x$ for $b^{2}=0$ and various $\beta $}
\label{fig6}
\end{figure}

In figure \ref{fig4}, the EoS parameter of HDE (\ref{1-9}) with
respect to $x $ is plotted for $\beta \approx 0.44$. It shows that
for all values of $b^{2} $ the EoS parameter behaves similar to $q$.
Likewise, from Eq. (\ref{1-11}), we find that the acceleration phase
is started from $w_{D}\leq -0.47$ and at present it gives
$w_{D}=-1.04$ which shows a phantom DE behavior.
The effective EoS parameter, which can determine the phases of acceleration
of the universe, plotted in Fig. \ref{fig5} for $\beta \approx 0.44$. It
shows that for all values of $b^{2}$ the universe transits from quintessence
($-0.34>w_{\mathrm{eff}}>-1$) to phantom phase ($w_{\mathrm{eff}}<-1$). From
this figure we see that at early time, roughly $x<-2$ which can be
translated to $z>6.4$, and hence $w_{\mathrm{eff}}\rightarrow 0$, which
indicates a pressureless DM dominated universe, for $b^{2}=0$. However, for $%
b^{2}>0$, the effective EoS parameter tends to finite negative
value. Therefore, in non-interacting case, HDE can give a unified
DM-DE profile in BD cosmology.
It must be noted that the parameter $\beta $ plays a crucial role in
$w_{\mathrm{eff}} $. This fact shows in Fig. \ref{fig6}. At present,
by decreasing $\beta <0.71 $, the universe tends to phantom phase in
such away that at $\beta =0.35$, phantom wall is crossed and bellow
this, the phantom phase is achieved. At late time it approaches to a
finite value, according to values of $\beta $.
At last we study on the squared of sound speed. As it is shown in Fig. \ref%
{fig9}, from now to past, $x\leq 0$, $v_s^2$ is negative for
$b^2\geq0.35$. Also increasing $\beta$, results a reduction of
$v_s^2$. It is worth noting that for $b^2<0.35$, and $\beta=0.44$,
we find $v_s^2>0$ and hence our model can lead to a stable DE
dominated universe favored by observations at the
present time. On the other hand, choosing values $\beta =0.44$ and $%
b^{2}>0.35$ leads to instable DE dominated universe. This implies
that with increasing $b^2$ we have more instability in the universe.

\begin{figure}[tbp]
\epsfxsize=7cm\centerline{\epsffile{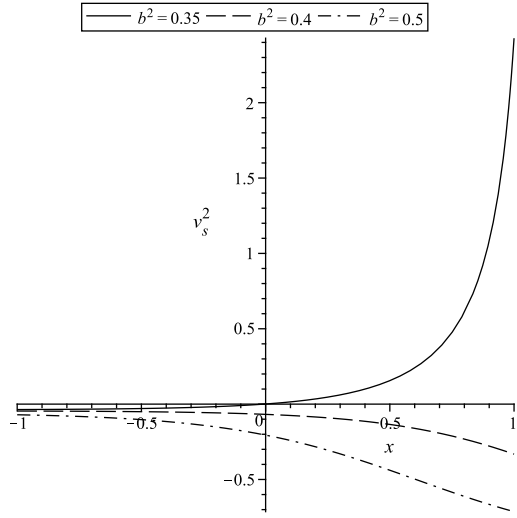}} \caption{Evolution
of $v_s^2$ versus $x$ for $\protect\beta=0.44$ and various $b^2 $
(flat universe)} \label{fig9}
\end{figure}

\section{CONCLUDING REMARKS}

\label{sum} In this paper, we studied interacting HDE model in the
framework of BD cosmology. As system's IR cutoff we chose GO cutoff
inspired by Ricci scalar curvature proposed by \cite{GO2008}. We
investigated cosmological implications of this model in ample
details. First, deriving the energy density ratio, $u(x)$ versus
$x=-\ln(1+z)$, showed that this ratio is bounded between finite
values of order unity according to interaction parameter so that by
increasing $b^{2}$, $u(x)$ evolves slower and will reach to a larger
finite values at infinity. It was also demonstrated that the cosmic
coincidence problem can only be alleviated in the interacting model
at late time. After that, we calculated the deceleration parameter
and the EoS parameter of HDE, effective EoS
parameter, and the squared of sound speed ($v_{s}^{2}$) in the case of $%
\gamma=2\beta$ (Ricci DE case). From the analysis of $q$, we found
that for the spatially flat FRW, the best value of $\beta$ is $\beta
\approx 0.44$. Present acceleration expansion, put an upper limit on
$\beta <0.71$ and for supper acceleration expansion, this limit
reduced to $\beta <0.36$. We found that the interaction affects on
the transition point of deceleration to acceleration. More
interaction moves the transition point to older universe.

The analysis of $w_{D}$ showed that the acceleration is started from
$ w_{D}\leq -0.47$ and at the present time the phantom DE phase with
$ w_{D}=-1.04$ can be achieved. We found that for both interacting
and non-interacting cases, the phantom divide of HDE can be crossed
with suitable choice of the parameter, $\beta =0.44$. Also the
analysis of $w_{ \mathrm{eff}}$ demonstrated that $\beta $ plays a
crucial role in phase of the evolution of the universe. At present,
by decreasing $\beta <0.71$, the universe tends to phantom phase in
such a way that at $\beta =0.36$, phantom wall is crossed and bellow
this the phantom phase is achieved. Considering interaction revealed
that for all values of $b^{2}$ the universe transits from
quintessence ($-0.34>w_{\mathrm{eff}}>-1$) to phantom phase ($w_{
\mathrm{eff}}<-1 $). Also, we showed that in non-interacting case,
HDE can give a unified DM-DE model in BD cosmology except that it
would not solve the coincidence problem at far future.

We also showed that in limiting case where
$\omega\rightarrow\infty$, our model reduces to the HDE model with
GO cutoff in standard cosmology \cite {GO2008}. We found that in the
framework of BD cosmology, crossing the phantom divide line for the
EoS parameter of HDE with GO cutoff can be more easily achieved for
than when resort to the Einstein gravity.

Finally, we investigate sound instability $v_s^2=\frac{dP}{d\rho}$
of the model. If $v_s^2$ is positive the HDE would be stable against
perturbations. When $v_s^2$ is negative we encounter the instability
in the background spacetime. Analyzing the speed of sound
$v_{s}^{2}$ indicates that HDE with GO cutoff in the framework of BD
cosmology can lead to a stable DE dominated universe favored by
observations, provided we take $\beta=0.44$ and $b^2<0.35 $. This is
in contrast to HDE model in Einstein gravity which does not lead to
a stable DE dominated universe \cite{myung1}.\\

\textbf{Acknowledgment}

The work of A. Sheykhi has been supported financially by Research
Institute for Astronomy and Astrophysics of Maragha (RIAAM), Iran.


\begin{thebibliography}{99}
\bibitem{RO1998} A. G. Riess et al., Astron. J. \textbf{116}, 1009 (1998)

\bibitem{WMAP9} G. Hinshaw et al. Astrophys. J. Suppl. 208 19,(2013)

\bibitem{HAN2000} S. Hanany et al., Astrophys. J. Lett. \textbf{545}, L5
(2000); C.B. Netterfield et al., Astrophys. J. \textbf{571}, 604 (2002);
D.N. Spergel et al., Astrophys. J. Suppl. \textbf{148}, 175 (2003).

\bibitem{COL2001} M. Colless et al., Mon. Not. R. Astron. Soc. \textbf{328},
1039 (2001); M. Tegmark et al., Phys. Rev. D \textbf{69}, 103501 (2004); S.
Cole et al., Mon. Not. R. Astron. Soc. \textbf{362}, 505 (2005); V.
Springel, C.S. Frenk, and S.M.D. White, Nature (London) \textbf{440}, 1137
(2006).

\bibitem{PADMAN2003} T. Padmanabhan, Phys. Rep. \textbf{380}, 235 (2003);
E. J. Copeland, M. Sami, S. Tsujikawa, Int. J. Mod. Phys. D \textbf{15},
1753 (2006).

\bibitem{Bisabr} Y. Bisabr, Phys. Rev. D \textbf{82}, 124041 (2010).

\bibitem{khodam11} A. Khodam-Mohammadi and M. Malekjani, Gen Relativ Gravit
\textbf{44}, 1163 (2012)

\bibitem{ZPAVON0606555} W. Zimdahl and D. pav\'{o}n, Class. Quant. Grav.
\textbf{24}, 5461 (2007).

\bibitem{SOLA2005} J. Sola and H. Stefancic, Phys. Lett. B \textbf{624}, 147
(2005); I.L. Shapiro, J. Sola, Phys. Lett. B \textbf{682}, 105 (2009).

\bibitem{HOOFT9310026} G. 't Hooft, [arXiv:gr-qc/9310026].

\bibitem{SUSS1995} L. Susskind, J. Math. Phys. (N.Y.) \textbf{36}, 6377
(1995).

\bibitem{COHEN1999} A. G. Cohen, D.B. Kaplan and A.E. Nelson, Phys. Rev.
Lett. \textbf{82}, 4971 (1999).

\bibitem{HSU2004} S. D. H. Hsu, Phys. Lett. B \textbf{594}, 13 (2004)
[arXiv:hep-th/0403052].

\bibitem{LI2004} M. Li, Phys. Lett. B \textbf{603}, 1 (2004)
[arXiv:hep-th/0403127].

\bibitem{MYUNG2007} Y. S. Myung, Phys. Lett. B \textbf{649}, 247(2007).

\bibitem{hsu} S. D. H. Hsu, Phys. Lett. B \textbf{669}, 275 (2008).

\bibitem{Pavon1} D. Pav\'{o}n, and W. Zimdahl, Phys. Lett. B \textbf{628},
206 (2005).

\bibitem{Sheykhi1} A. Sheykhi, Class.Quant.Grav.\textbf{27}, 025007 (2010).

\bibitem{GAO0712.1394} C. Gao, X. Chen and Y. G. Shen, arXiv:0712.1394
[astro-ph].

\bibitem{FENG2008} C. Feng, arXiv:0806.0673 [hep-th].

\bibitem{GO2008} L.N. Granda, and A. Oliveros, Phys. Lett. B \textbf{669}, 275 (2008).

\bibitem{easson2011} D. A. Easson, P. H. Frampton, G. F. Smoot,Phys.Lett.B \textbf{696}, 273 (2011).

\bibitem{GO2009} L.N. Granda, and A. Oliveros, Phys. Lett. B \textbf{671}, 199 (2009), arXiv:0810.3663 [gr-qc].

\bibitem{khodam111} A. Khodam-Mohammadi, Mod. Phys. Lett. A \textbf{26}, 2487 (2011).

\bibitem{m-kh2011} M. Malekjani, A. Khodam-Mohammadi,Int.J.Mod.Phys.D \textbf{20}, 281 (2011).

\bibitem{Sharif} M. Sharif, Abdul Jawad, Eur. Phys. J. C \textbf{72}, 2097
(2012).

\bibitem{Malek} M. Malekjani, A. Khodam-Mohammadi, N. Nazari-pooya, Astrophys. Space Sci. \textbf{332}, 515
(2011).

\bibitem{Li} Miao Li, Xiao-Dong Li, Jun Meng, Zhenhui Zhang, Phys. Rev. D \textbf{88}, 023503
(2013).

\bibitem{bas-pol2012}S. Basilakos, D. Polarski, J. Sola,Phys.Rev. D \textbf{86}, 043010 (2012),arXiv:1204.4806 [gr-qc].

\bibitem{bas-sola2014} S. Basilakos, J. Sola,Phys. Rev. D \textbf{90}, 023008 (2014),arXiv:1402.6594 [astro-ph].

\bibitem{JORDAN1955} P. Jordan, , Nature \textbf{164}, 637 (1955), Schwerkraft und Weltall (Friedrich Vieweg
und Sohn, Braunschwig)

\bibitem{BD1961} C. Brans, and R.H. Dicke, Phys. Rev. \textbf{124}, 925
(1961).

\bibitem{GONG2004} Y. Gong, Phys. Rev. D \textbf{70} (2004) 064029; H. Kim,
H.W. Lee, Y.S. Myung, Phys. Lett. B \textbf{628},11 (2005); A. Sheykhi,
Phys. Lett. B \textbf{681},205 (2009); B. Bertotti, L. Iess, P. Tortora,
Nature \textbf{425}, 374 (2003); V. Acquaviva, L. Verde, JCAP \textbf{12},
001 (2007); L. Xu, J. Lu, W. Li, arXiv:0905.4174 [astro-ph.CO]; M.R. Setare,
M. Jamil, Phys. Lett. B \textbf{690},1 (2010).

\bibitem{SHEYKHI0907.5458} A. Sheykhi, Phys. Lett. B \textbf{681}, 205
(2009).

\bibitem{XULI2009} L.X. Xu, W.B. Li, and J.B. Lu, Eur. Phys. J. C \textbf{60}
135 (2009).

\bibitem{JAMILKARAMI2011} M. Jamil et al. Int. J. Theor. Phys, \textbf{51}
,604 (2012).

\bibitem{BP2007} N. Banerjee, and D. Pavon, Phys. Lett. B \textbf{647}, 447
(2007).

\bibitem{RONG-GEN} R-G. Cai, and Q. Su, Phys. Rev. D \textbf{81}, 103514
(2010).

\bibitem{Planck} Jun-Qing Xia, JCAP \textbf{11}, 022 (2013).

\bibitem{peebleratra} P. J. E. Peebles and B. Ratra, Rev. Mod. Phys.
\textbf{75} 559 (2003).

\bibitem{Wein-book} S. Weinberg, \textit{Gravitation and Cosmology},
John Wiley \& Sons, Inc. (1972).

\bibitem{pavon2001} N. Banerjee, and D., Pavon, Class. Quant. Grav. \textbf{%
18}, 593 (2001).

\bibitem{XULU2011} J. Lu, W. Wang, L. Xu, Y. Wu, Eur. Phys. J. Plus  \textbf{126}, 92 (2011), arXiv:1105.1868v3 [astro-ph].

\bibitem{LU-liu2012} J. Lu, L. Ma, M. Liu, Y. Wu,International Journal of Modern Physics D \textbf{21}, 1250005
(2012),arXiv:1203.4906v1 [astro-ph] .

\bibitem{PD2012} D. Pavon et al., Phys. Rev. D \textbf{86}, 083509 (2012).

\bibitem{PLANCK} Planck Collaboration, arXiv:1303.5076;\newline
J. Q. Xia, H. Li, X. Zhang, Phys. Rev. D 88, 063501 (2013)
%Dark Energy Constraints after Planck%

\bibitem{SUKUZI2012} N. Suzuki et al., Astrophys. J. (ApJ), \textbf{746}, 85
(2012).

\bibitem{XIALI2013} J-O. Xia, H. Li, and X. Zhang, Phys. Rev. D \textbf{88},
063501 (2013).

\bibitem{ray2007}  S. Ray and U. Mukhopadhyay, Int. J. Mod. Phys. D \textbf{16}, 1791 (2007).

\bibitem{j-sola2008} J. Sola. J. Phys. A \textbf{41}, 164066 (2008), arXiv:0710.4151v2 [hep-th].

\bibitem{grande-sola2008} J. Grande, J. Sola et al, arXiv:1001.0259v2 [astro-ph.CO].

\bibitem{myung1} Y. S. Myung, Phys. Lett. B 652  223 (2007).

\bibitem{myung2} K. Y. Kim, H. W. Lee and Y. S. Myung, Phys. Lett. B 660
(2008) 118.

\bibitem{gorini} V. Gorini, A. Kamenshchik, U. Moschella, V. Pasquier and A.
Starobinsky, Phys. Rev. D 72 (2005) 103518.

\bibitem{sandvik} H. Sandvik, M. Tegmark, M. Zaldarriaga and I. Waga, Phys.
Rev. D 69 (2004) 123524.

\bibitem{ebrins} E. Ebrahimi and A. Sheykhi, Int. J. Mod. Phys. D \textbf{20} (2011)
2369;\newline E. Ebrahimi and A. Sheykhi, Int. J. Theor. Phys. 52
(2013) 2966.

\bibitem{saaidi} Kh. Saaidi, arXiv: 1202.4097.

\bibitem{WANGXU2010} Y. Wang, and L. Xu, Phys. Rev. D \textbf{81}, 083523
(2010).

\bibitem{gra-sola2006} J. Grande, J. Sola, H. Stefancic, JCAP 0608, 011 (2006).

\bibitem{gra-sola2009} J. Grande, A. Pelinson, J. Sola, Phys Rev D \textbf{79} 043006 (2009).

\end{thebibliography}
\end{document}